\documentclass[a4paper,11pt]{article}
\pdfoutput=1 % if your are submitting a pdflatex (i.e. if you have images in pdf, png or jpg format)

\usepackage{jinstpub} % for details on the use of the package, please see the JINST-author-manual

\title{\boldmath THE TRISTAN DETECTOR - \\[0.5ex] 2018-2019 LATITUDE SURVEY OF COSMIC RAYS}

\author[a,\dagger]{J.P. Saraiva,\note{Corresponding author.}}    %jinstpub.sty alterado para poder usar footnote com dagger e nao com número = 1 (senão o número 1 é dp repetido na 1ª footnote adicionada no documento).
\author[a]{A. Blanco,}
\author[b]{J. A. Garz\'{o}n,}
\author[b]{D. Garc\'{i}a-Castro,}
\author[a]{L. Lopes}
\author[c]{and V. Villasante-Marcos,}

\affiliation[a]{LIP, Laborat\'{o}rio de Instrumenta\c{c}\~{a}o e F\'{i}sica Experimental de Part\'{i}culas,\\Rua Larga, Physics Department, 3004-516 Coimbra, Portugal}
\affiliation[b]{LabCAF, IGFAE. Universidade de Santiago de Compostela,\\Santiago de Compostela, Spain}
\affiliation[c]{Instituto Geogr\'{a}fico Nacional,\\Madrid, Spain}

\emailAdd{joao.saraiva@coimbra.lip.pt}

\abstract{In 2018-2019 a cosmic-ray latitude survey at sea level was performed by the TRISTAN detector, an autonomous system composed by three planes of RPCs (120 x 150 cm$^{2}$). The detector made a two-way journey on board of the Spanish vessel Sarmiento de Gamboa between Vigo (Spain) and Punta Arenas (Chile), measuring continuously the cosmic-ray rate throughout the Atlantic crossing. In this work, we present the results of the first journey, correlating the obtained variation of the cosmic-ray rate with the vertical cutoff rigidity, as well as presenting some details of the detector and its autonomous DAQ system used during the campaign.}

\keywords{resistive-plate chambers}

\collaboration[c]{on behalf of the LIP - RPC group}

\proceeding{XV$^{\text{th}}$ Workshop on Resistive Plate Chambers and Related Detectors\\
  10-14 February 2020\\
  Rome}

\begin{document}
\maketitle
\flushbottom

\section{Introduction}
\label{sec:intro}

\begin{figure}[htbp]
\centering
\includegraphics[width=.9\textwidth, trim = 0 0 0 0, clip=true, angle=0]{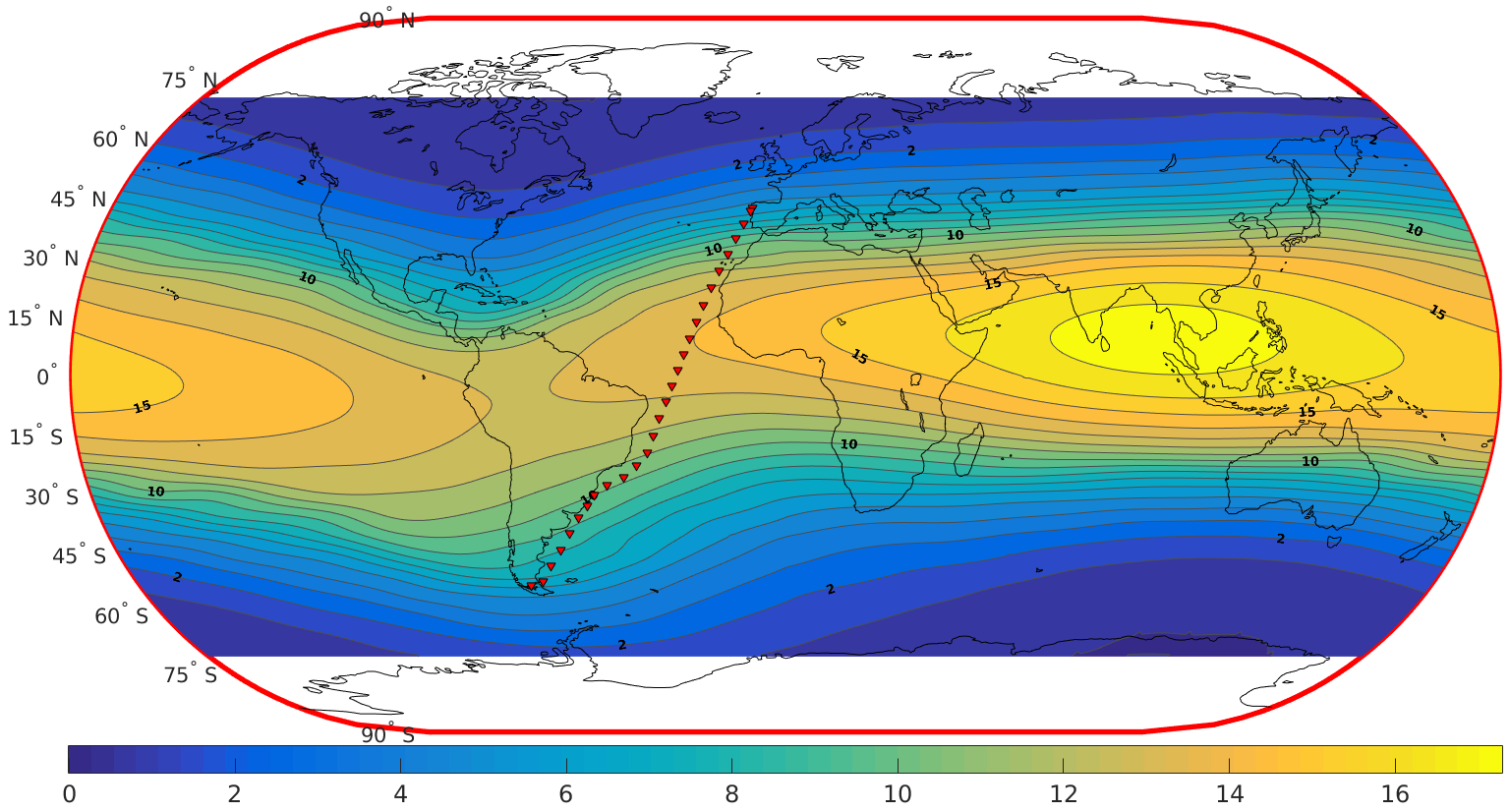}
\caption{Vertical cosmic-ray cutoff rigidity [GV] with iso-rigidity contours (Epoch = 1980); path covered by the boat Sarmiento de Gamboa during the cosmic-ray latitude survey of the first journey (one red triangle per day).}
\label{cutOff}
\end{figure}

It is well known that primary cosmic rays interact with the Earth's atmosphere, producing secondary cosmic rays that, in turn, can reach the ground and be detected at sea level. A minimum of energy is required for these primary particles to overcome the magnetosphere and reach the atmosphere. This energy threshold depends on the geographic location, being higher close to the equatorial plane. A map of cutoff energies, calculated for particles hitting the atmosphere vertically and expressed by the so called vertical cutoff rigidity, is shown in figure~\ref{cutOff}.

The vertical cutoff rigidity, the minimum rigidity\footnote{Momentum per unit charge (pc/Ze) [V] (e.g. a 10 GeV proton has a rigidity of 10 GV). Calculated by Shea and Smart [1983].} a particle must have to cross the geomagnetic field and reach a given geographic location, was interpolated from tabulated values available here ~\cite{chapt6-ref}. Figure~\ref{cutOff} also shows the iso-rigidity contours, as well as the path covered by the boat Sarmiento de Gamboa during the cosmic-ray latitude survey of 2018-2019, as described later.

Cosmic-ray latitude surveys carried out over the years e.g. by boat, plane or balloon allow us to directly correlate the variation of the cosmic-ray rate with the cutoff rigidity. Complementary to these surveys, neutron monitors installed all over the world measure the nucleonic component of secondary cosmic rays ~\cite{nmdb-ref}.

On November 2018, the Spanish vessel Sarmiento de Gamboa started a latitude survey from Vigo (Spain, 42$^{\circ}$N) to the Spanish Antarctic base Juan Carlos I (62$^{\circ}$S), with several detectors measuring continuously the rate of secondary cosmic rays throughout the Atlantic crossing.
As part of the Antarctic Cosmic-Ray Observatory (ORCA) ~\cite{blanco-ref}, the TRISTAN detector, described in this paper, 
made a two-way journey in the boat, from November 2018 to April 2019, however only as far as Punta Arenas (Chile, 53$^{\circ}$S).

\section{TRISTAN Detector}
\label{sec:tristan}

Built in Coimbra with the expertise of the mechanical workshop and detector laboratory teams of LIP~\cite{lip-ref}, including the HV power supplies and the gas system with monitoring capability, the TRISTAN detector consists of three Resistive-Plate Chambers (RPC), each of them with an active area of 120 x 150 cm$^{2}$.

\begin{figure}[ht]
\centering
\includegraphics[width=.76\textwidth, trim = 0 0 0 0, clip=true, angle=0]{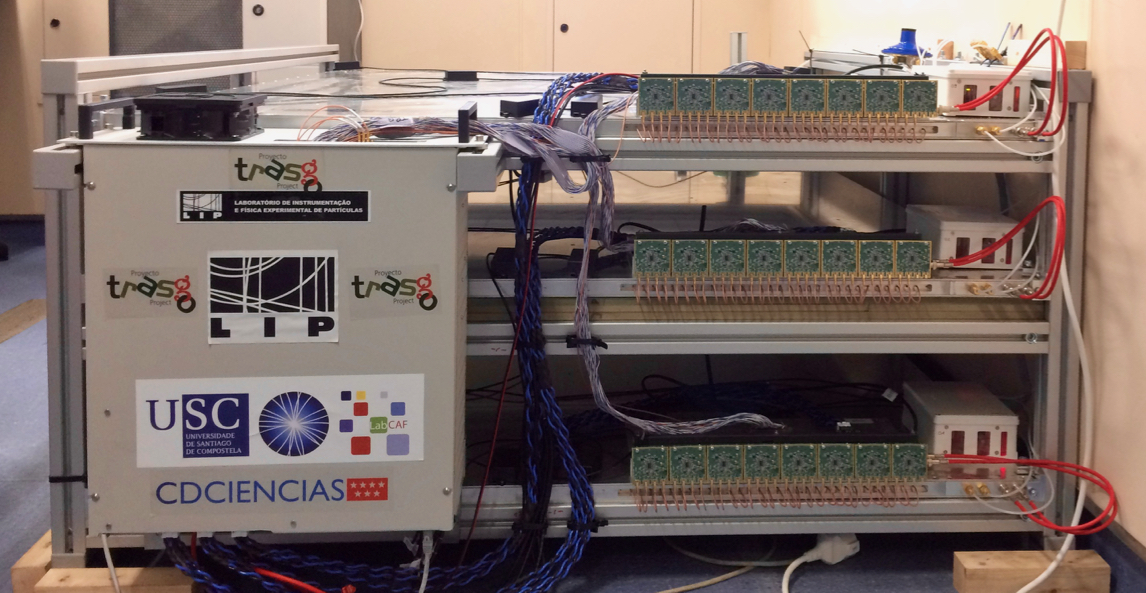}
\caption{The TRISTAN detector in the Sarmiento de Gamboa.}
\label{rpcInboat}
\end{figure}

Figure~\ref{rpcInboat} shows the detector with the three layers of RPCs installed in the boat Sarmiento de Gamboa, in a room with controlled temperature. TRISTAN is a multi-track detector working in avalanche mode, being operated in open gas loop with pure freon 134a and a gas flow rate of 11 cc/min in each plane.

\begin{figure}[ht]
\centering
\includegraphics[width=1.0\textwidth, trim = 0 0 0 0, clip=true, angle=0]{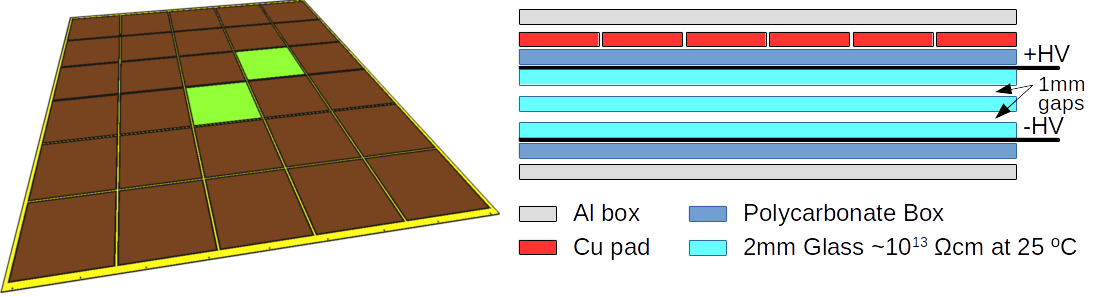}
\caption{Left: 6 x 5 pads per plane with guard rings 6 mm wide in-between them; right: schematic of the different RPC constituent layers.}
\label{rpcDetalhes}
\end{figure}

Each plane is read out by 30 pads 24 x 25 cm$^{2}$, with a design based on the R\&D done in the Auger framework~\cite{auger-ref}. A total of 90 readout channels instrumented with a front-end electronics measuring time and charge in each single channel with a timing resolution around 35 ps~\cite{FEE-ref}. To avoid crosstalk between pads, grounded guard rings 6 mm wide separate them, as shown in  figure~\ref{rpcDetalhes}. Each plane of RPCs has two gaps 1 mm wide separated by 2 mm thick glass (see figure~\ref{rpcDetalhes}).

The DAQ system of the TRISTAN detector is fully autonomous:
\begin{itemize}
\item data acquisition starting automatically after reboot;
\item in case of communication failure, established procedures were followed e.g. power-cycling the respective device (readout board, i2c system, router, etc.);
\item the log analysis of cosmic rates, environmental sensors, gas flow rates, HV, etc., was done every 30 minutes, searching for values out of range;
\item alarms were sent via email in case of issue;
\item daily generated reports, with preliminary data analysis, were also sent via email;
\item real-time data such as cosmic rates \& coincidences were showed in a screen at the entrance of the physics department in Coimbra and updated every 30 minutes to avoid bandwidth saturation.
\end{itemize}

A total of 90 channels were read out by the TRB3 readout board via 4 FPGA-based TDCs with a timing resolution around 20 ps~\cite{trb3-ref}. A quad-core single board computer Odroid-C2 was used to process the data and communicate with all the other devices. In case of computer crash, a watchdog could reboot the Odroid after 15 minutes. Communication with environmental sensors (temperature, pressure and relative humidity), gas monitoring sensors (pressure and flow rate) and the HV power supplies was done via i2c protocol.

The trigger scheme of TRISTAN is based on the coincidence between signals coming from the RPCs \#1 and \#3 (starting counting from the bottom) with the coincidence time window set to 30 ns.

In order to maintain the detector gain constant during the whole period of the survey, the provided HV was automatically adjusted as a function of pressure and temperature according to \eqref{HV}~\cite{auger-ref}.

\begin{equation} \label{HV}
HV [V] = \frac{E/N[Td] \cdot d_{gap}[cm] \cdot P[mbar]}{0.01138068748 \cdot T[K]},
\end{equation}

with $E/N$ the reduced electric field, $d_{gap}$ the gap thickness, $P$ and $T$ the pressure and temperature, respectively.

\section{Results}
\label{sec:results}

\begin{figure}[ht]
\centering
\includegraphics[width=0.7\textwidth, trim = 0 0 0 0, clip=true, angle=0]{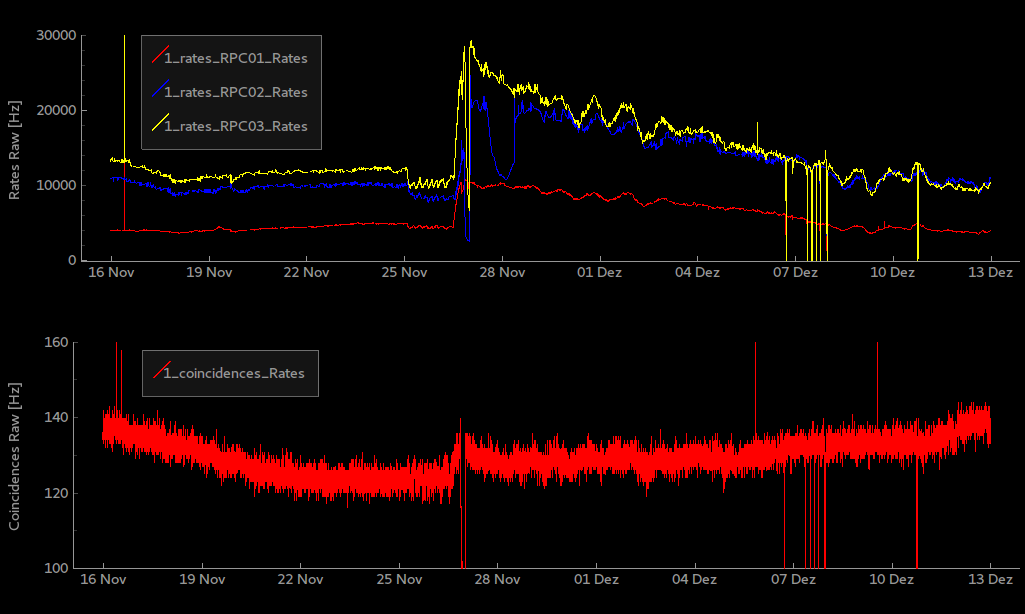}    %RawRatesInvertColors.png
\caption{Top: background rates of the RPCs along the survey; bottom: raw coincidences between the RPC \#1 and the \#3 (counting from the bottom to the top).}
\label{rawRates}
\end{figure}

As shown in figure~\ref{rawRates}, the background rates of the RPC planes during the first days of the survey oscillated around an average value of 10 kHz ({\raise.17ex\hbox{$\scriptstyle\mathtt{\sim}$}}1 Hz/cm$^{2}$). Then, a significant increase of the rates occurred on November 26th due to a failure of the air conditioning (hereinafter AC-failure) when the boat was nearby the Equator. According to the environmental sensors installed around the detector, the temperature increased from 20 to 30 $^{\circ}$C in about 10 hours. Afterwards, the rates decreased over the following days while the boat was approaching the Antarctica. The same figure shows the raw coincidences between the RPCs \#1 and \#3, giving us already an overview of how the cosmic-ray flux evolved along the survey. However, the detector was clearly affected by the increase in temperature due to the AC-failure, and some corrections had to be applied to overcome this issue, as described later.

\begin{figure}[htbp]
\centering
\includegraphics[width=0.49\textwidth, trim = 0 0 0 0, clip=true, angle=0]{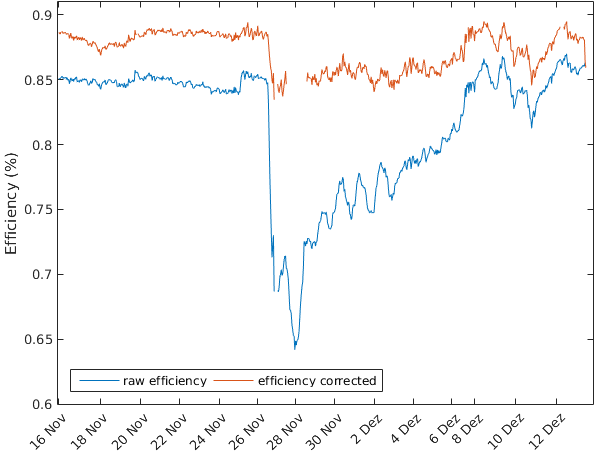}
\quad
\includegraphics[width=0.47\textwidth, trim = 0 0 0 0, clip=true, angle=0]{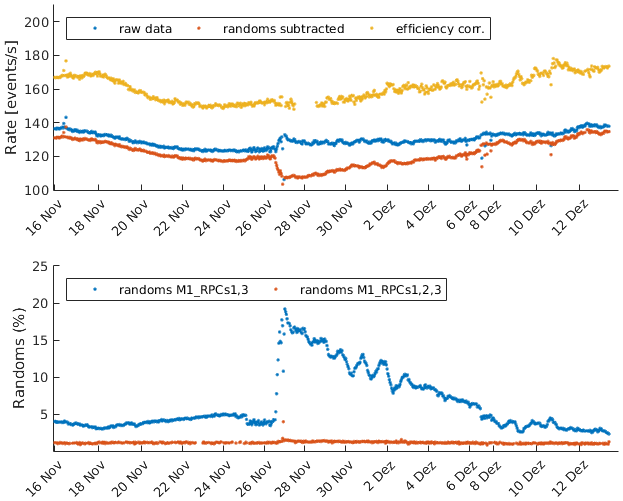}
\caption{Left: efficiency of the RPC \#2 computed with and without randoms; right-top: coincidences between RPCs \#1 and \#3 (raw, corrected for randoms and for randoms \& efficiency); right-bottom: estimated random coincidences via polynomial fitting, requiring hit multiplicity equal to one in two and three RPCs.}
\label{efficiency-rates}
\end{figure}

The efficiency, calculated for the layer that was not in the trigger (RPC \#2), with and without randoms, is shown in figure~\ref{efficiency-rates}. To compute the efficiency, events with only one pad with signal (multiplicity of one, M1) in both RPCs \#1 and \#3 were compared with the events seen by the RPC \#2. As it can be seen from the mentioned figure, even correcting for random coincidences, the efficiency decreased few percent with the AC-failure and this despite the HV adjustment previously mentioned.

The efficiency obtained for the RPC \#2 is slightly below 90\% because of the guard rings between pads, which area corresponds to {\raise.17ex\hbox{$\scriptstyle\mathtt{\sim}$}}10\% of the total RPC active area. Moreover, the lack of data in the curve of the corrected efficiency is owed to the significant increase in temperature and respective current saturation of the HV power supply in the second plane during more and less one day.

\begin{figure}[htbp]
\centering
\includegraphics[width=0.55\textwidth, trim = 0 0 0 0, clip=true, angle=0]{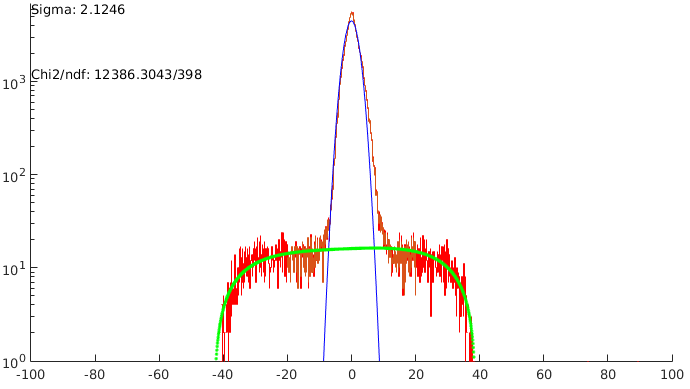}
%\quad
\includegraphics[width=0.44\textwidth, trim = 0 0 0 0, clip=true, angle=0]{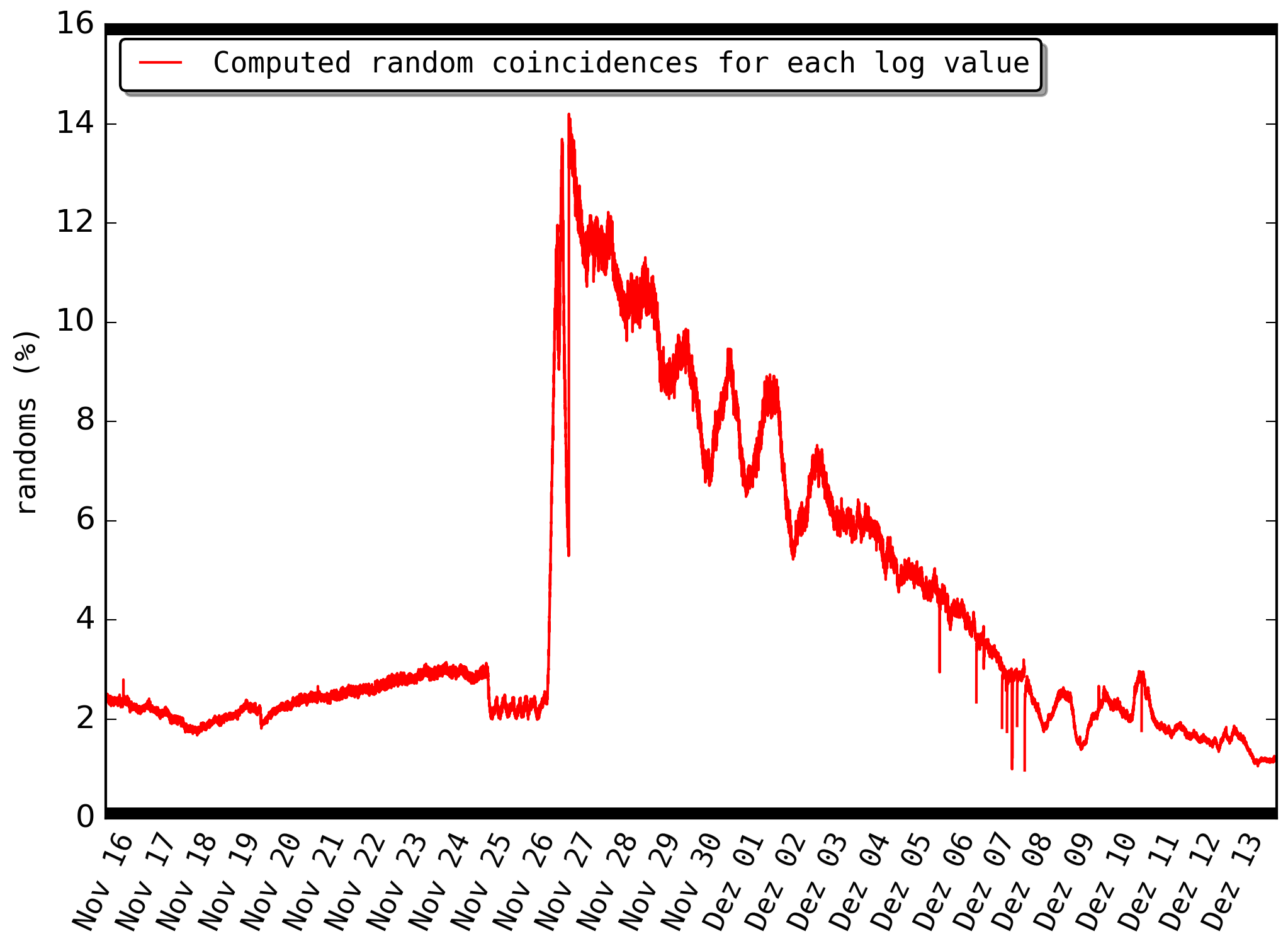}
\caption{Left: example showing the subtraction between the time of events from the RPCs in the trigger (\# of events vs. subtracted time [ns]). The central peak, related to true coincidences of particles that crossed both planes in more and less 1ns, have first to be removed with a Gaussian fit (here at 4 sigma; blue curve). The two regions left in both sides of the Gaussian, which correspond to random coincidences, are then estimated with a polynomial function as shown by the green curve; right: computed percentage of random coincidences using \eqref{randoms}.}
\label{polynomialFit-calculatedRandoms}
\end{figure}

Figure~\ref{efficiency-rates} also shows three curves of the cosmic-ray coincidences between planes \#1 and \#3: without any correction (as previously showed in figure~\ref{rawRates}), correcting for random coincidences and correcting for both - randoms and efficiency. Only when correcting for both, the randoms and the efficiency, it was possible to mitigate the effect of the AC-failure, keeping the cosmic-ray rate constant before and after the failure. Just as in the case of the corrected efficiency, the lack of data in the curve of the corrected secondary cosmic-ray rate is due to the current saturation of the HV power supply in the RPC \#2. In the same figure, the random coincidences can be seen, estimated using a fourth-order polynomial fit (see also figure~\ref{polynomialFit-calculatedRandoms}) and requiring M1 in two and in three RPCs.

The random coincidences can also be calculated using \eqref{randoms}~\cite{knoll-ref}. By doing so, one could see a difference of {\raise.17ex\hbox{$\scriptstyle\mathtt{\sim}$}}2-3\% between estimated and calculated values of the percentage of random coincidences when requiring M1 in the RPCs \#1 and \#3 (with the estimated values systematically above the calculated ones; see figures ~\ref{efficiency-rates} and ~\ref{polynomialFit-calculatedRandoms}).

\begin{equation} \label{randoms}
R = 2 \cdot r_{RPC1} \cdot r_{RPC3} \cdot T,
\end{equation}

with $R$ the rate of random coincidences, $r_{RPC1}$ and $r_{RPC3}$ respectively the background rates of planes \#1 \& \#3 and $T$ the coincidence time window (set to 30 ns).

The same exercise was carried out requiring M1 in the three RPCs and using \eqref{randoms2}~\cite{knoll-ref}.

\begin{equation} \label{randoms2}
R = 3 \cdot r_{RPC1} \cdot r_{RPC2} \cdot r_{RPC3} \cdot T^{2},
\end{equation}

In this case, the polynomial fit of the randoms (also showed in figure~\ref{efficiency-rates}) overestimated the respective analytical values, and will therefore be improved. In fact a triple coincidence decreases significantly the random coincidences making their estimation much more difficult.

\begin{figure}[htbp]
\centering
\includegraphics[width=0.44\textwidth, trim = 0 0 0 0, clip=true, angle=0]{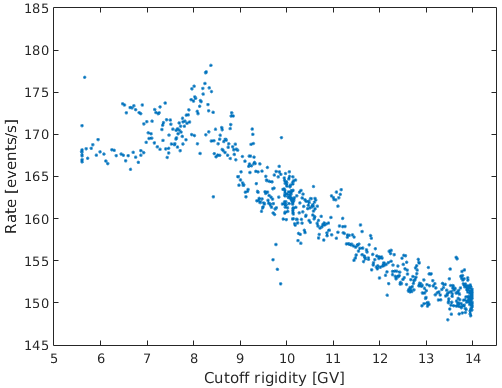}
\quad
\includegraphics[width=0.525\textwidth, trim = 0 0 0 0, clip=true, angle=0]{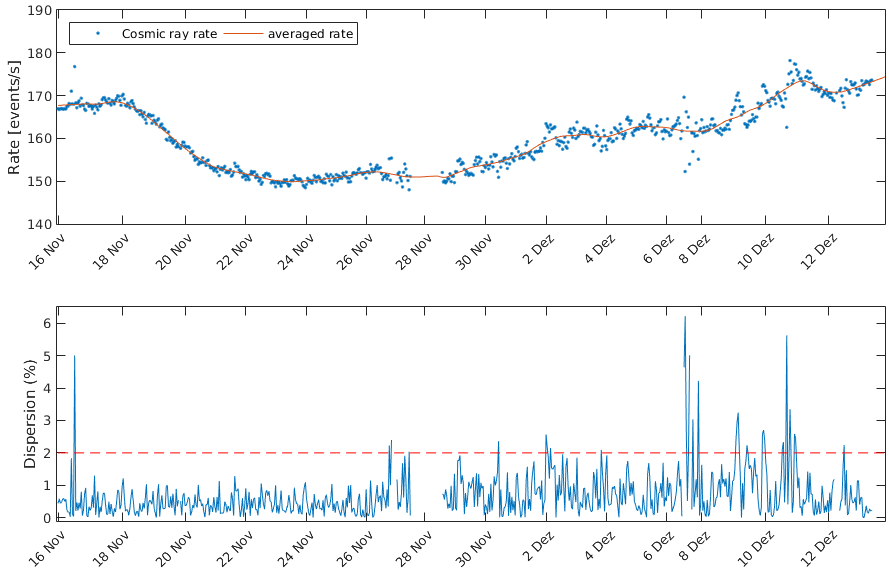}
\caption{Left: secondary cosmic-ray rate vs. vertical cutoff rigidity of primary cosmic rays.; right: dispersion of the cosmic-ray rate around a 1.5-day moving average.}
\label{ratesRigidity-dispersion}
\end{figure}

The variation of the corrected secondary cosmic-ray rate obtained with the TRISTAN detector during the latitude survey can now be compared with the respective values of the cutoff rigidity of primary cosmic rays already showed in figure~\ref{cutOff}. This is shown in figure~\ref{ratesRigidity-dispersion}, where the correlation between both quantities is clearly evidenced. Some points obtained at the beginning of the latitude survey seems to be slightly outside the correlation region and are currently being investigated.

The corrected secondary cosmic-ray rate can also be compared with a moving average rate over {\raise.17ex\hbox{$\scriptstyle\mathtt{\sim}$}}1.5 days. As shown in figure~\ref{ratesRigidity-dispersion}, almost all the points are below a dispersion of 2\% from the average rate. This result could even be improved if the temperature was controlled during the whole journey (a significant increase of the dispersion can be seen in the figure after the AC-failure).

The low dispersion of cosmic-ray measurements evidenced by the TRISTAN detector allow us to conclude that it could be used to measure Forbush Decreases\footnote{Decrease of few/several percent in galactic cosmic rays on Earth due to the increase in solar cosmic rays (in turn related to the increase in solar activity).} which are relevant to study the Sun's activity.

\section{Conclusions}
\label{sec:conclusions}

The TRISTAN detector, with three RPCs designed and built in Coimbra, made a latitude survey along the Atlantic Ocean from November 2018 to April 2019, measuring successfully the cosmic-ray flux. The detector is currently (2020) installed in the Spanish Antarctic station Juan Carlos I (Livingston Island - Antarctic Peninsula). The DAQ system is fully autonomous and was used with success during the whole campaign.

Despite a failure of the air conditioning in the room where the detector was located, the correlation between the secondary cosmic-ray rate and the vertical cutoff rigidity of primary cosmic rays was clearly evidenced.

The detector efficiency decreased few \% with the 10 $^{\circ}$C variation and this despite the HV adjustment as a function of the pressure and temperature.

The dispersion of the measurements of the secondary cosmic-ray rate showed to be below 2\% with respect to a 1.5-day moving average. A value that could be improved with a controlled room temperature during the whole survey. TRISTAN could therefore be used for precise measurements of the cosmic-ray flux and Forbush Decreases, relevant for Solar Physics.

\end{document}